\documentclass{elsart}

\usepackage{amssymb}

\usepackage{amsmath}
\newtheorem{theorem}{Theorem}[section]
\newtheorem{corollary}[theorem]{Corollary}
\newtheorem{lemma}[theorem]{Lemma}

\newtheorem{tab}[theorem]{Table}

\begin{document}

\begin{frontmatter}

\title{A doubling construction for self-orthogonal codes}

\author{Vladimir D. Tonchev
}


\address{
Department of Mathematical Sciences, Michigan Technological University,
Houghton, Michigan 49931, USA, tonchev$@$mtu.edu}

\vspace{5mm}

\begin{abstract}
A simple construction of quaternary hermitian self-orthogonal codes with parameters
$[2n+1,k+1]$ and $[2n+2,k+2]$ from a given pair of self-orthogonal
$[n,k]$ codes, and its link to quantum codes is considered.
As an application, an optimal quaternary linear $[28,20,6]$ dual containing
code is found that yields a new optimal $[[28,12,6]]$ quantum code.

\end{abstract}

\begin{keyword}
  hermitian self-orthogonal code \sep quantum code 
\MSC   94B. 
\end{keyword}
\end{frontmatter}

\section{Introduction}

We assume familiarity with the basics of classical error-correcting codes
\cite{McS}.

The {\em hermitian} inner product in $GF(4)^n$ is defined as
\begin{equation}
\label{eq2}
(x,y)_H =  \sum_{i=1}^n  x_{i}y_{i}^{2},
\end{equation}
while the {\em trace } inner product in $GF(4)^n$ is defined as
\begin{equation}
\label{eq3}
(x,y)_T = \sum_{i=1}^n (x_{i}y_{i}^2 + x_{i}^{2}y_i).
\end{equation}
A code $C$ is {\em self-orthogonal} if $C\subseteq C^{\perp}$,
and {\em self-dual} if $C=C^{\perp}$.
A linear code $C\subseteq GF(4)^n$ is self-orthogonal with respect to the
trace product (\ref{eq3}) if and only if it is self-orthogonal with respect 
to the hermitian product (\ref{eq2}) \cite{CRSS}.

An {\em additive} $(n,2^k)$ code $C$ over $GF(4)$ is a subset of $GF(4)^n$
consisting of $2^k$ vectors which is closed under addition.
An additive code is {\em even} if the weight of every codeword is even, and
otherwise {\em odd}. Note that an even additive code is 
trace self-orthogonal,
and a linear self-orthogonal code is even \cite{CRSS}.
If $C$ is an $(n,2^k)$ additive code with weight enumerator
\begin{equation}
\label{id1}
 W(x,y)=\sum_{j=0}^n A_{j}x^{n-j}y^j, \end{equation}
the weight enumerator of the trace-dual code $C^{\perp}$ is given by 
\begin{equation}
\label{id2}
W^{\perp}= 2^{-k}W(x+3y,x-y) \end{equation}
In \cite{CRSS}, Calderbank, Rains, Shor and Sloane described a method
for the construction of quantum error-correcting codes from additive 
codes that are self-orthogonal with respect to the trace product 
(\ref{eq3}).
\begin{thm}
\cite{CRSS}
\label{th1}
An additive trace self-orthogonal $(n,2^{n-k})$ code $C$ such 
that there are no vectors of weight $< d$ in $C^{\perp}\setminus C$ 
yields 
a quantum code with parameters $[[n,k,d]]$.
\end{thm}
A quantum code  associated with an additive code $C$ is {\em pure} if 
there are no vectors of weight $<d$ in $C^{\perp}$; otherwise, the code
is called {\em impure}. A quantum code is called {\em linear}
if the associated additive code $C$ is linear.

A table with lower and upper bounds on the minimum distance $d$ for 
quantum $[[n,k,d]]$ codes
of length $n\le 30$ is given in the paper by Calderbank, Rains, Shor and Sloane 
\cite{CRSS}.  An extended version of this table was compiled
by Grassl \cite{Gr}. 
Bounds on the minimum distance of linear codes 
are available on line at  http://www.codetables.de.

\section{A doubling construction}

\begin{lemma}
\label{lem1}
Suppose that $C_i$ ($i=1, 2$) is a linear hermitian self-orthogonal $[n,k]$ code over $GF(4)$ with
generator matrix $G_i$, and $x^{(i)} \in C_{i}^{\perp}$ is a vector of odd weight.\\

(a)  The code
$C'$ with  generator matrix
\begin{equation}
\label{G'}
G'=\left( \begin{array}{ccc|ccc|c}
 &&&&&& 0 \\
 & G_1 && & G_2 && \ldots  \\
&&& &&& 0\\
\hline
& x^{(1)} && 0 & \ldots & 0 & 1 
\end{array}\right)
\end{equation}
is a hermitian self-orthogonal self-orthogonal $[2n+1,k+1]$ code with dual distance
\begin{equation}
\label{dC'}
 d(C')^{\perp} \le {\rm min}(d(C_{11}^{\perp}), d(C_2^{\perp})),
\end{equation}
where $C_{11}$ is the code spanned by the rows of $G_{11}$
given by (\ref{g11}):
\begin{equation}
\label{g11}
G_{11}=\left( \begin{array}{c|c}
 & 0\\
 G_1 & \dots\\
 & 0\\
\hline
 x^{(1)} & 1
\end{array}\right).
\end{equation} 
(b) The code $C''$  with  generator matrix
\begin{equation}
\label{G''}
G''=\left( \begin{array}{ccc|ccc|cc}
 &&&&&& 0 & 0\\
 & G_1 && & G_2 && \ldots & \\
&&& &&& 0 & 0\\
\hline
& x^{(1)} && 0 & \ldots & 0 & 1 & 0\\
\hline
0 & \ldots & 0 & & x^{(2)} & & 0 & 1
\end{array}\right)
\end{equation}
is a hermitian self-orthogonal self-orthogonal $[2n+2,k+2]$ code with dual distance
\begin{equation}
\label{dC''}
 d(C'')^{\perp} \le {\rm min}(d(C_{11}^{\perp}), d(C_{22}^{\perp})),
\end{equation}
where  $C_{22}$ is the code spanned by the rows of $G_{22}$ given by (\ref{g22}):
\begin{equation}
\label{g22}
G_{22}=\left( \begin{array}{c|c}
 & 0\\
 G_2 & \dots\\
 & 0\\
\hline
 x^{(2)} & 1
\end{array}\right).
\end{equation}
 
\end{lemma}

{\bf Proof}.
The self-orthogonality of $C'$ and $C''$ follows from the fact that all rows of $G'$ and $G''$
have even weights, and every two rows of $G'$, as well as every two rows of $G''$,
are pairwise orthogonal.
Since the weight of $x^{(1)}$ (resp. $x^{(2)}$) is odd,  $x^{(1)}$ does not belong to $C_1$,
and  $x^{(2)}$ does not belong to $C_2$, and that implies the dimensions
of $C'$ and $C''$. The bounds (\ref{dC'}), (\ref{dC''})  on the dual distance
follow trivially by the observation that every codeword of $C_{11}^\perp$
(resp. $C_{22}^\perp$) extends to a codeword of $(C')^\perp$ (resp $(C'')^\perp$)
by filling in all remaining coordinates with zeros.
\qed

Using the connection to quantum codes described in Theorem \ref{th1},  
Lemma \ref{lem1}  implies the following.

\begin{corollary}
\label{cor}
The existence of quaternary hermitian self-orthogonal $[n,k]$ codes $C_i$ ($i=1, 2$)
satisfying the assumptions of Lemma \ref{lem1} implies
the existence of a pure quantum linear  $[[2n+1,2n-2k-1,d']]$
code with $d' \le {\rm min}(d(C_{11}^{\perp}), d(C_2^{\perp}))$, 
 and a pure quantum linear  $[[2n+2,2n-2k-2,d'']]$ code 
with $d'' \le {\rm min}(d(C_{11}^{\perp}), d(C_{22}^{\perp})) $.
\end{corollary}

We will apply Lemma \ref{lem1} and its Corollary \ref{cor} to some self-orthogonal
codes of length $n=2k+1$ being shortened codes of self-dual $[2k+2,k+1]$ codes.

For example, the matrix
\[ G_1 =\left( \begin{array}{ccccc} 1 & 0 & 1 & \alpha & \alpha\\ 0 & 1 & \alpha & \alpha & 1
\end{array}\right) \]
is the generator matrix of a self-orthogonal $[5,2,4]$ code $C_1$ over $GF(4)=\{ 0, 1, \alpha, \alpha^2 \}$.
The code $C_1$ is a shortened code of the unique (up to equivalence) self-dual $[6,3,4]$ code.
Applying Lemma \ref{lem1} with $C_2 = C_1$, $G_2 = G_1$, and $x^{(1)} = x^{(2)}$ being the 
all-one vector of length 5,
gives a self-orthogonal $[11,3]$ code $C'$ with dual distance 3 and a self-orthogonal $[12,4]$ code $C''$
with dual distance 4, which give  optimal quantum $[[11,5,3]]$ and $[[12,4,4]]$ codes respectively 
via Corollary \ref{cor}.

Similarly, a pair of self-orthogonal $[7,3]$ codes obtained as shortened codes of the
unique (up to equivalence) self-dual $[8,4,4]$ code can be used to obtain optimal
quantum $[[15,7,3]]$ and $[[16,6,4]]$ codes.

The smallest parameters of a self-dual quaternary linear code that yields a quantum code with minimum
distance $d\ge 5$ via Corollary \ref{cor} are $[14,7,6]$. The only such code, 
up to equivalence,
is the quaternary extended quadratic residue code $q_{14}$ \cite[page 340]{NRS}.
We apply Lemma \ref{lem1} using the pair of self-orthogonal $[13,6]$ codes
$C_1$, $C_2$  generated by the following matrices:
\[ G_1=\left(\begin{array}{c}
 0 0 0 0 1 0 0 2 1 0 2 3 3\\
 3 0 0 0 0 1 0 0 2 1 0 2 3\\
 3 3 0 0 0 0 1 0 0 2 1 0 2\\
 2 3 3 0 0 0 0 1 0 0 2 1 0\\
 0 2 3 3 0 0 0 0 1 0 0 2 1\\
 1 0 2 3 3 0 0 0 0 1 0 0 2
\end{array}\right), \ G_2=\left(\begin{array}{c}
 0 0 0 0 1 1 3 0 2 3 0 0 2\\
 2 0 0 0 0 1 1 3 0 2 3 0 0\\
 0 2 0 0 0 0 1 1 3 0 2 3 0\\
 0 0 2 0 0 0 0 1 1 3 0 2 3\\
 3 0 0 2 0 0 0 0 1 1 3 0 2\\
 2 3 0 0 2 0 0 0 0 1 1 3 0
 \end{array}\right), \]
where for convenience, the elements $\alpha$ and $\alpha^2$ of $GF(4)$
are written as 2 and 3 respectively. The matrices $G_1$, $G_2$ are circulant.
The codes $C_1$, $C_2$ are cyclic and  equivalent to a shortened code of $q_{14}$.

Choosing $x^{(1)}=x^{(2)}$ to be the all-one vector of length 13, we obtain
the generator matrix $G'$ (\ref{G'}) of a self-orthogonal $[27,7]$ code $C'$ with dual distance 5,
and the generator matrix $G''$ (\ref{G''}) of a self-orthogonal $[28,8]$ code with dual 
distance 6. The matrix $G''$ is available on line at

\begin{verbatim}
http://www.math.mtu.edu/~tonchev/gm28-8.html
\end{verbatim}

By Corollary \ref{cor}, $C'$ gives a pure optimal quantum $[[27,13,5]]$ code, while $C''$ gives
a pure optimal quantum $[[28,12,6]]$ code. 

An alternative geometric construction of a quantum code with the first parameters, $[[27,13,5]]$, 
was given by the author in \cite{ton}. To the best of our knowledge, the quantum code with the second 
parameters, $[[28,12,6]]$, is new (a quantum $[[28,12,5]]$ code was listed in \cite{CRSS}).

The weight distribution of the $[28,8]$ code $C''$ is given in Table \ref{tab1}.
\begin{tab}
\label{tab1}
\end{tab}
\begin{tabular}{|r|r|}
\hline
$w$ & $A_w$ \\
\hline
          12 &          39\\
          14 &           6\\
          16&        3198 \\
          18&        9204\\
          20&       18213\\
          22&       22854\\
          24&       10569\\
          26&        1248\\
          28&         204\\
\hline
\end{tabular}

The weight enumerator of the dual $[28,20]$ code $(C'')^\perp$ is
\[ 1 + 6240y^6 + 37128y^7 + 314223y^8 + 2044848y^9 + 11883768y^{10}+ \ldots \] 
We note that the code  $(C'')^\perp$ is an optimal linear $[28,8,6]$ quaternary code:
6 is the largest possible minimum distance of a quaternary linear $[28,8]$ code \cite{Gr}.

\section{Acknowledgments}

Magma \cite{Magma} was used  for some of the computations.
The author wishes to thank the unknown referee for noticing some typos
and discrepancies that were subsequently corrected. 


\end{document}